\documentclass[a4paper,11pt]{article}
\usepackage{graphicx}
\usepackage{jheppub}
\usepackage[T1]{fontenc} 
\title{\boldmath
Killing quantum entanglement by acceleration or a black hole} \author[a]{Yue Dai}
\author[a]{Zhejun Shen}
\author[a,b,1]{Yu Shi,\note{Corresponding author.}}
\affiliation[a]{Center for Field Theory and Particle Physics, Department of Physics  \&  State Key Laboratory of Surface Physics,   Fudan University,\\ Shanghai 200433, China}
\affiliation[b]{Collaborative Innovation Center of Advanced Microstructures, Fudan University, \\Shanghai 200433, China} \emailAdd{dy1983@gmail.com}
\emailAdd{victorshenzj@gmail.com}
\emailAdd{yushi@fudan.edu.cn}

\abstract{
We  consider two entangled accelerating qubits coupled with  real scalar fields, each described by the Unruh-Wald model. It is  demonstrated that because of the Unruh effect of the fields, the  bipartite  entanglement  between the two qubits suddenly dies when the acceleration of one or more qubits are large enough. We also consider three entangled accelerating qubits in GHZ state and in W state,  with equal acceleration-frequency ratio, and found that in either state, the tripartite entanglement suddenly dies at a certain value of acceleration-frequency ratio. The equivalence between the Rindler metric and the Schwarzchild metric in the vicinity of the horizon of a black hole implies that for two entangled qubits outside a black hole, the entanglement suddenly dies when one or both of the qubits are close enough to the horizon, while the three entangled qubits in GHZ or W state, the tripartite entanglement suddenly dies when these qubits are close enough to the horizon.}

\begin{document}
Published: JHEP {\bf 09} (2015) 071
\maketitle
\flushbottom

\section{Introduction}

Recent two decades witnessed both intensive and extensive investigations on quantum entanglement, ``the characteristic trait of quantum mechanics'' in the words of Schr\"{o}dinger~\cite{schroedinger}.  More recently, this trend has been extended to the realms of high energy physics. The relativistic effects of acceleration and gravitation on quantum entanglement were investigated, shedding new light on the subject of quantum effects of gravity.

A particularly concerned subject is  the Unruh effect, that is, the particle content of a field is observer-dependent, and thus an accelerating detector in the Minkowski vacuum of a field feels a thermal bath of particles of this field~\cite{davies,fulling,dewitt,unruh,unruhwald,matsas,mukhanov}.  The consequences of the Unruh effect on various kinds of entanglement have been studied. For example, because of the Unruh effect, a state of two field modes that is   maximally entangled in an inertial frame becomes less entangled when one of the modes is observed by an accelerating detector, and degrades with the increase of acceleration,  towards zero at the limit of infinite acceleration~\cite{fuentes}.  Analogous  entanglement degradation  occurs for field modes observed by two detectors close to the horizon of a black hole, with one of them freely falling while the other barely escapes~\cite{fuentes,eduardo}.  This problem was then extended to the case of three entangled modes, and it was found that  when one of  the modes is observed by an accelerating detector, the tripartite entanglement, which cannot be reduced to all kinds of bipartite entanglement, does not approach zero in the infinite acceleration limit~\cite{hwang,shamirzai}.  Extension of such studies to Fermion fields was also made~\cite{alsing,jing1,jing2,richter}. These results are consistent with the general feature that entanglement of a  field state depends on the choice of single particle modes~\cite{shi}.  Entanglement between detectors have also been studied in the case that one of the two entangled detectors accelerates while the other moves uniformly, and was found to exhibit entanglement sudden death~\cite{yu}. One approach is to model  the detectors  as  harmonic oscillators~\cite{hu0}, for which calculations were also made on the physical details~\cite{hu1,hu2}.  Another approach is to model the detectors as  qubits~\cite{matsas1}. Yet another approach is to consider the detectors as an open quantum system~\cite{doukas1,doukas2,huyu}.  Other quantum informational quantities such as discord~\cite{datta,celeri} and quantum Fisher information~\cite{yao,wang} have also been studied. So far there was no work on the case that more than one of the entangled detectors accelerate.

In this paper,  we consider two causally separated but quantum-entangled qubits, each of which independently accelerates and is coupled with a scalar field as described by the Unruh-Wald model. Our work originated in generalizing a previous work on the decoherence of one qubit due to acceleration~\cite{kok}.  We show that because of  the Unruh effect of the fields, the entanglement between the qubits vanishes at finite values of acceleration instead of in the infinite limit,  i.e. exhibit entanglement sudden death. Our result implies that outside of the black hole, the entanglement between the two entangled detecting qubits vanishes when one or both of the qubits are close enough to the horizon. We also report a result on the  tripartite entanglement in three entangled accelerating qubits.

\section{Formalism}

Let us consider two qubits A and B far away from each other, that is, we assume there is no causal connection between the two qubits.  For each qubit,  we apply the model of the Unruh and Wald~\cite{unruhwald}.

The Hamiltionian of each qubit q (q=A,B) is
\begin{equation}
H_q=\Omega_q Q_q^\dagger Q_q, \label{qubit}
\end{equation}
where the creation operator $Q_q^\dagger$ and annihilation operator $Q_q$ are defined by
$Q_q | 0 \rangle_q = Q_q^\dagger  | 1 \rangle_q = 0,$ $
Q_q^\dagger  | 0 \rangle_q =  | 1 \rangle_q$ and
$Q_q | 1 \rangle_q = | 0\rangle_q$, with subscript q=a,b. $\Omega_q$  gives the energy difference between eigenstates $|1\rangle_q$ and $|0\rangle_q$.

Each qubit is locally coupled with a field $\Phi_q$ within a small region around it,   the interaction Hamiltonian being
\begin{equation}
H_{I_q}(t_q) = \epsilon_q  ( t_q  ) \int_{\Sigma_q}  {\Phi_q  (  \mathbf{x}  ) [ {\psi_q  ( \mathbf{ x}  )Q_q  + {\psi_q^*} ( \mathbf{x} ){Q_q^\dag }}  ]\sqrt { - g} {d^3}x}.
\end{equation}
where $\mathbf{x}$ and $t_q$ are spacetime coordinates in the comoving frame of the qubit,  the integral is over the spacelike Cauchy surface $\Sigma_q$ at given time $t_q$, $\epsilon_q  ( t_q  )$ is the coupling constant with a finite duration of qubit-field interaction, $\psi_q (\mathbf{ x} )$ is a smooth function nonvanishing within a small volume around the qubit~\cite{unruh}. The fields $\Phi_A$ and $\Phi_B$ could be the same or different.

We presume that the distance between the two qubits is so large  that there is no physical coupling or influence between the neighboring fields of the two qubits during the  interaction times.  The total Hamiltonian is simply $H_A+H_{\Phi_A}+ H_{I_A}+H_B+H_{\Phi_B}+H_{I_B}$, where $H_{\Phi_q}$ is the Klein-Gordon Hamiltonian for $\Phi_q$. In the Minkowski spacetime, each qubit is confined in its own Rindler wedge and possesses boost Killing fields which are timelike. The only extra constraint on the trajectories of the two qubits is that the time interval of the interaction between each qubit and its neighboring field   multiplied by the speed of light is smaller than  the shortest distance between the  interaction regions of the two qubits.

Therefore after a time duration longer than the interacting times $T_q \gg 1/\Omega_q$, the state of the whole system in the interaction picture is transformed  by  $$U_{A}\otimes U_{B},$$
where $U_{q}$ is the unitary transformation acting on qubit  $q$ and the field  $\Phi_q$ in its neighboring region, as given by the Unruh-Wald model. To the first order~\cite{unruhwald},
\begin{equation}
U_{q} \approx 1-i \int \Phi_{q0}(t',\mathbf{x})\epsilon_q(t') [Q_{q0}e^{-i\Omega_q t'} \psi_q(\mathbf{x}) + Q_{q0}^\dagger e^{i\Omega_q t'} \psi_q^*(\mathbf{x})]\sqrt{-g'} d^3x dt',
\end{equation}
where $\Phi_{q0}$ and $Q_{q0}$ are $\Phi_q$ and $Q_q$ for $\epsilon_q=0$, i.e. when the qubit-field coupling is turned off. It can be obtained that
\begin{equation}
U_{q} \approx 1+iQ_{q0} a^\dagger (\Gamma_q^*)  -  i  Q_{q0}^\dagger a(\Gamma_q^*),
\end{equation}
where $a(\Gamma_q^*)$  and $a^\dagger(\Gamma_q^*)$ are the  annihilation and  the creation operators of  $\Gamma_q^*$, with
\begin{equation}
\Gamma_q (x) \equiv -2i \int [G_R(x;x')-G_A(x;x')]\epsilon_q(t')e^{i\Omega t}\psi_q^*(\mathbf{x}')\sqrt{-g'}
d^4x',
\end{equation}
$G_{Rq}$ and $G_{Aq}$ being the retarded and advanced Green functions of the field $\Phi_q$, respectively.

For any mode $\chi_q$, we can write
\begin{eqnarray}
a(\Gamma_q^*) & = &\langle \Gamma_q^*,\chi_q\rangle   a(\chi_q) +  \langle \Gamma_q^*,\chi'_q\rangle a(\chi'_q),
\\
a^\dagger (\Gamma_q^*) &= & \langle \Gamma_q^*,\chi_q \rangle^*   a^\dagger (\chi_q) +  \langle \Gamma_q^*,\chi'_q\rangle^*  a^\dagger (\chi'_q),
\end{eqnarray}
where $\chi'_q$ is some mode orthogonal to $\chi_q$. $\langle \Gamma_q^*,\chi_q\rangle =  \frac{i}{2}\int_{\Sigma_q}  [\Gamma_q\partial_\mu \chi_q
-(\partial_\mu\Gamma_q)\chi_q]dS^\mu$, where $\Sigma_q$ is some Cauchy surface.  This inner product can be assumed to be negligible unless $\chi_q$ is at a frequency $\approx \Omega_q$. Therefore, each qubit q is only coupled with the field mode $\chi(\Omega_{q})$    with frequency $\Omega_q$, which is further assumed to be nondegenerate.      Hence we only need to consider $\chi_{\Omega_{A}} $ and $\chi_{\Omega_{B}}$ in studying the qubits A and B. All the other modes are decoupled with the qubits.

Now we consider the Fock state $|n\rangle_{\Omega_q}$, containing $n$ particles in the mode $\chi(\Omega_{q})$ of the field $\Phi_q$, as observed in the Rindler wedge confining qubit q,
\begin{equation}
\begin{array}{rcccl}
a(\Gamma_q^*) |n\rangle_{\Omega_q} & =& \mu_q  a(\Omega_q) | n\rangle_{\Omega_q} &=& \mu_q  \sqrt{n} | n-1\rangle_{\Omega_q}, \\
a^\dagger (\Gamma_q^*) |n\rangle_{\Omega_q}&=&\mu_q^* a^\dagger (\Omega_q) | n\rangle_{\Omega_q}& =& \mu_q^*  \sqrt{n+1} | n+1\rangle_{\Omega_q},
\end{array}
\end{equation}
where  $\mu_q  \equiv \langle \Gamma_q^*,\chi_{\Omega_q}\rangle
=  \int \epsilon_q(t)  e^{i\Omega_q t}\psi_q^*(\mathbf{x}) \chi_{\Omega_q}(t,\mathbf{x})\sqrt{-g} d^4x$.

Hence $U_q$ evolves only the qubit $q$ and the mode $\chi_{\Omega_q}$, while the other modes of $\Phi_q$ are not affected and can be ignored,
\begin{eqnarray}
U_{\Omega_q} | 0 \rangle_q  | n \rangle_{\Omega_q}&= & | 0 \rangle_q | n \rangle_{\Omega_q}  - i\sqrt n \mu_q | 1 \rangle_q  | {n - 1} \rangle_{\Omega_q}, \label{s1}  \\
U_{\Omega_q} | 1 \rangle_q | n \rangle_{\Omega_q} & =  & | 1 \rangle_q | n \rangle_{\Omega_q}
+ i\sqrt {n + 1} \mu_q^* | 0 \rangle_q | {n + 1} \rangle_{\Omega_q}.  \label{s2}
\end{eqnarray}

\section{Entangled states of the two detecting qubits}

We now suppose the initial state of the two qubits to be
\begin{equation}\label{ints}
|\Psi_i\rangle = \alpha {| 0 \rangle _A}   {| 1 \rangle _B} + \beta {| 1 \rangle _A} {| 0 \rangle _B},
\end{equation}
where $\alpha$ and $\beta$ are superposition coefficients satisfying $|\alpha|^2+|\beta|^2=1$.  The results for the initial state of the form of
$\alpha {| 0 \rangle _A}   {| 0 \rangle _B} + \beta {| 1 \rangle _A} {| 1 \rangle _B}$ are similar.  Without causal connection between the two qubits or between the fields, each qubit detects a thermal bath of the Unruh particles  determined by its own acceleration. With each qubit in its own Rindler wedge,  the initial state of the whole system, as observed by the observers comoving with the qubits, is described by the density matrix
\begin{equation}
\rho_i = |\Psi_i\rangle\langle \Psi_i|\otimes \rho_{\Omega_A}\otimes \rho_{\Omega_B} \otimes \rho',
\end{equation}
where
\begin{equation}
\rho_{\Omega_q}=C_q\sum_{n_q} e^{ - 2\pi n_q \Omega_q/a_q}|n_q\rangle_{\Omega_{q}}\langle n_q|,
\end{equation}
$a_q$ is the acceleration of qubit q, $C_q \equiv \sqrt {1 - {e^{ - 2\pi \Omega_q/a_q}}}$, $\rho'$ is the state of the other decoupled modes and is ignored henceforth.

The final state of the system in the interaction picture  is
\begin{equation}
\rho_f =U_{B}^\dagger U_{A}^\dagger \rho_i U_{A}U_{B},
\end{equation}
which can be evaluated by substituting  Eqs.(\ref{s1}) and (\ref{s2}). Subsequently by tracing out the fields, we obtain the reduced density matrix of the two qubits, with respect to the comoving observers

\begin{equation}
\begin{array}{rl}
\rho_{AB} = & C_A^2C_B^2\sum\limits_{n_A,n_B} \frac{ e^{ - 2\pi  ( n_A\Omega _A / a_A  +  n_B  \Omega _B / a_B  ) } }{ Z_{ n_A , n_B } } \nonumber\\ & \times \{
 [  | \alpha  | ^2    (   n_B  + 1   )|\mu _B|^2  +  | \beta  | ^2    (   n_A  + 1   )|\mu _A|^2 ]|  00  \rangle \langle 00| \nonumber\\  & + \alpha  \beta ^* |  01  \rangle \langle 10| + \beta  \alpha ^* |  10  \rangle \langle 01|  \nonumber\\ & +  [ | \alpha  |^2  + | \beta  |^2  n_B  ( n_A + 1   )|\mu _B|^2|\mu _A|^2  ]|  01  \rangle \langle 01|  \nonumber\\
 &  +  [| \alpha  |^2  n_A (n_B + 1)|\mu _A|^2 |\mu _B|^2 + | \beta  |^2 ]| 10 \rangle \langle 10|
 \nonumber\\ & +  [ | \alpha  |^2   n_A |\mu _A|^2 +  | \beta  |^2   n_B |\mu _B|^2  ]| 11 \rangle \langle 11|  \},
\end{array}
\end{equation}
where $
{Z_{{n_A},{n_B}}} \equiv  1 + {n_A}|\mu _A|^2 +  {{n_B} |\mu _B|^2}  +  {n_A} {n_B}  |\mu _A|^2|\mu _B|^2  +|\alpha|^2 (|\mu_B|^2+ n_A|\mu_A|^2|\mu_B|^2) + | \beta  |^2 ( |\mu _A|^2 +  {n_B} |\mu _B|^2|\mu _A|^2 ).$

In the case that qubit A moves uniformly while qubit B accelerates, $\rho_{AB}$ is
\begin{equation}
\begin{array}{rl}
\rho_{AB} = & C_B^2\sum\limits_{n_B} \frac{ e^{ - 2\pi   n_B  \Omega _B / a_B  } }{ Z_{ n_B } } \nonumber\\ & \times \{
 [  | \alpha  | ^2    (   n_B  + 1   )|\mu _B|^2  +  | \beta  | ^2    |\mu _A|^2 ]|  00  \rangle \langle 00| \nonumber\\  & + \alpha  \beta ^* |  01  \rangle \langle 10| + \beta  \alpha ^* |  10  \rangle \langle 01|  \nonumber\\ & +  [ | \alpha  |^2  + | \beta  |^2  n_B  |\mu _B|^2|\mu _A|^2  ]|  01  \rangle \langle 01|  \nonumber\\
 &  +   | \beta  |^2 | 10 \rangle \langle 10| +    | \beta  |^2   n_B |\mu _B|^2  | 11 \rangle \langle 11|  \},
\end{array}
\end{equation}
where $
{Z_{n_B}} \equiv 1 +  {n_B} |\mu _B|^2 +|\alpha|^2 |\mu_B|^2 + | \beta  |^2 |\mu _A|^2+ |\beta|^2 {n_B} |\mu_A|^2  |\mu _B|^2.$

We now study the correlation and entanglement in $\rho_{AB}$. Note that the entanglement and correlation are respectively the same in Schr\"{o}dinger and interaction pictures.

\section{ von Neumann entropy  $S ( {{\rho_{AB}}}  )$  and mutual information $I ( {A:B}  )$  }

The von Neumann entropy
\begin{equation}
S ( {{\rho_{AB}}}  )\equiv -\mathrm{Tr}\rho_{AB}\log\rho_{AB}
\end{equation} is a measure of mixture  of $\rho_{AB}$. On the other hand, in the Minkowski frame, the state of the whole system is a pure state, and $S ( {{\rho_{AB}}} )$ quantifies the entanglement between the two qubits on one hand, and the fields on the other.

From $S(\rho_{AB})$, we also calculate the mutual information
\begin{equation}
I ( {A:B}  ) \equiv  S ( {{\rho _{A}}}  ) + S ( {{\rho _{B}}}  ) - S ( {{\rho _{AB}}}  ),
\end{equation}
where \begin{equation}\rho_{A} = \mathrm{Tr}_B\rho_{AB},
\end{equation}
\begin{equation} \rho_{B} = \mathrm{Tr}_A\rho_{AB}.
\end{equation}

$I ( {A:B}  )$ is the difference between the sum of the entropies of $A$ and $B$ as a whole  on one hand, and the entropy of $A$ plus $B$ as a whole on the other, and  is thus a quantification of the total correlation contributed by both entanglement and classical correlation.  In the numerical calculations throughout this paper, the bases of the logarithms are chosen to be $2$, and the parameters $\mu_A$, $\mu_B$ are both  set to be $0.1$.

\begin{figure}
\includegraphics[width=3.2in]{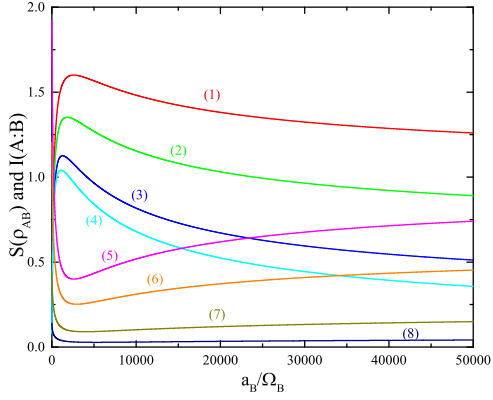}
\caption{\label{figqf1} $S(\rho_{AB})$ [plots (1) to (4)] and  mutual information $I(A:B)$ [plots (5) to (8)] as  functions of the acceleration-frequency ratio of qubit  B, in the case that qubit A moves uniformly, for different initial states. (1,5): $\alpha=1/\sqrt{2}$, (2,6): $\alpha=0.4$, (3,7): $\alpha=0.2$, (4,8): $\alpha=0.1$.  }
\end{figure}

\begin{figure}
\includegraphics[width=3.2in]{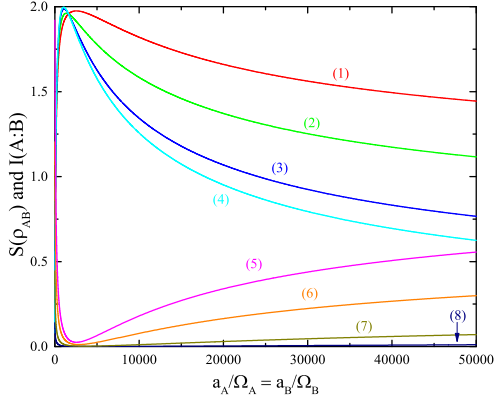}
\caption{\label{figqf2} $S ( {{\rho_{AB}}}  )$ [plots (1) to (4)] and  mutual information $I(A:B)$ [plots (5) to (8)] as  functions of the acceleration-frequency ratio, which is   assumed to be the same for the two qubits, for   different initial states. (1,5): $\alpha=1/\sqrt{2}$, (2,6): $\alpha=0.4$, (3,7): $\alpha=0.2$, (4,8): $\alpha=0.1$. The extreme values are   insensitive to the initial state.
}
\end{figure}

\begin{figure}
\includegraphics[width=3.2in]{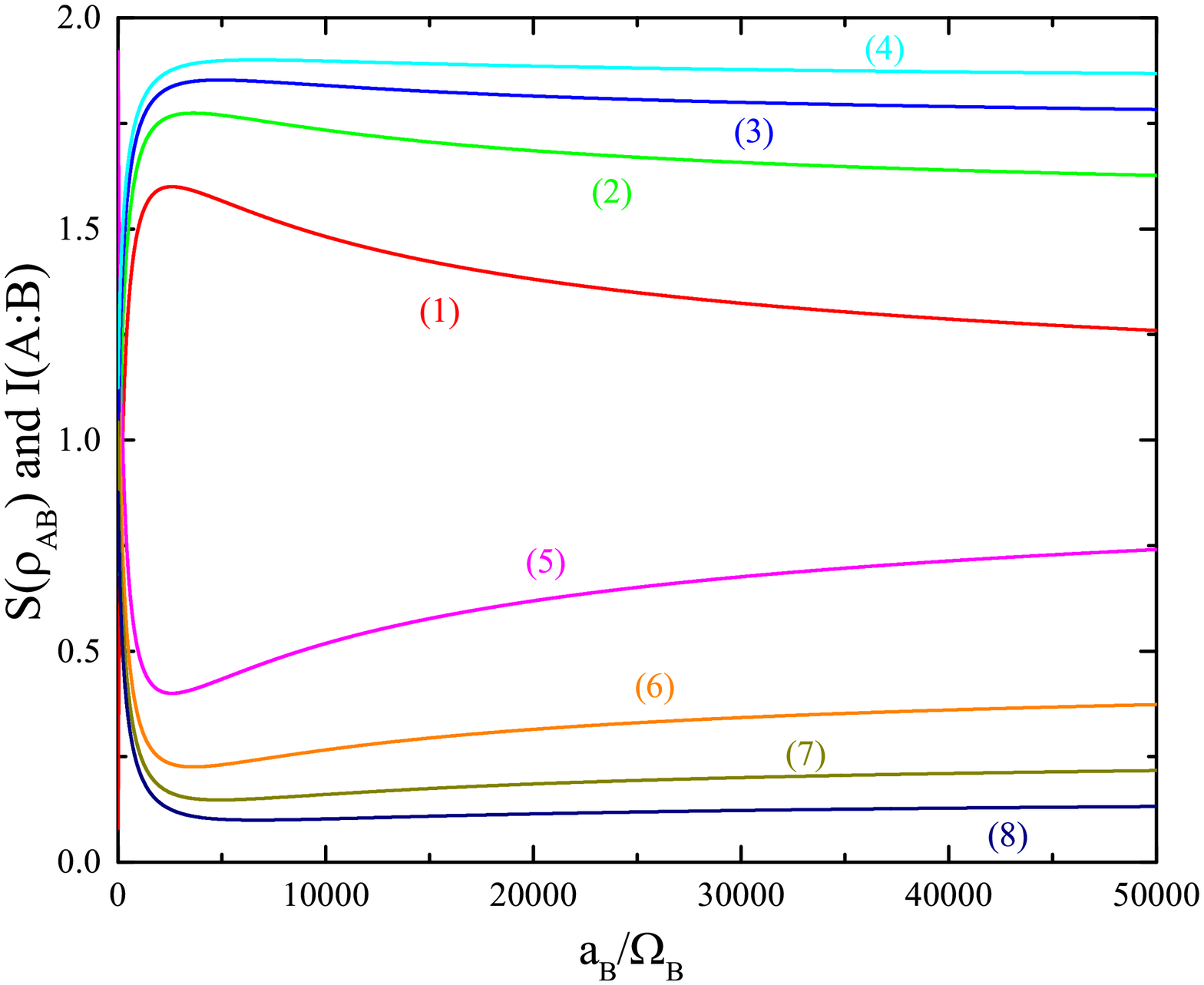}
\caption{\label{figqf3} $S ( {{\rho_{AB}}}  )$  [plots (1) to (4)] and  mutual information $I(A:B)$ [plots (5) to (8)] as  functions  of the acceleration-frequency ratio of qubit $B$, for different given values of acceleration-frequency ratio  of qubit $A$:  (1,5): $a_A/\Omega_A=0$, (2,6):$a_A/\Omega_A=100$, (3,7): $a_A/\Omega_A=200$, (4,8): $a_A/\Omega_A=300$.   It is set that $\alpha=\beta=\frac{1}{\sqrt{2}}$.}
\end{figure}

$S ( {{\rho_{AB}}}  )$  and $I ( {A:B}  )$  are  depicted together in  Figs.~\ref{figqf1} to \ref{figqf3} for three cases.  The result for the case of qubit A uniformly moving while qubit B accelerating is shown in Fig.~\ref{figqf1}. The special case of two maximally entangled qubits with one of them uniformly moving was previously studied by using a different approach and assuming no coupling between the uniformly moving qubit and the field~\cite{matsas}. In our studies, both qubits always couple with the fields.
Fig.~\ref{figqf2} depicts the result for the case that the acceleration-frequency ratios of the two qubits are always equal.  Fig.~\ref{figqf3} gives the results  for various given values of $a_A/\Omega_A$. As shown in these figures, when both  $a_A/\Omega_A$ and $a_B/\Omega_B$ are close to $0$, $\rho_{AB}$ is close to the pure state $|\Psi_i\rangle$, hence  $\rho_A$ is  close to $|\alpha|^2|0\rangle \langle 0|+|\beta|^2 |1\rangle\langle 1|$ while   $\rho_B$ is  close to $|\beta|^2|0\rangle \langle 0|+|\alpha|^2 |1\rangle\langle 1|$ , therefore $S ( {{\rho_{AB}}}  )$ is close to $0$ while $I(A:B)$ is close to $-2|\alpha|^2\log |\alpha|^2-2|\beta|^2\log |\beta|^2$. With the increase of one or both of the acceleration-frequency ratios,  $S ( {{\rho_{AB}}} )$  quickly increases up to a maximum while $I ( {A:B}  )$  quickly decreases down to a minimum. The actual extreme values depend on the details of the dynamics, but we can make the following estimation.  If $\rho_{AB}$ is maximally mixed, $S(\rho_{AB})$ reaches the absolute  maximum $2$ while $S(\rho_A)$ and $S(\rho_B)$ reach the absolute maximum $1$,  because $\rho_{AB}$ is $4$-dimensional while $\rho_A$ and $\rho_B$ are $2$-dimensional. Therefore the minimal value of $I(A:B)$ is near $0$.    When the acceleration-frequencies further increase, $S ( {{\rho_{AB}}}  )$ slowly decreases while  $I(A:B)$ slowly increases. In the limit of $a_B/\Omega_B \rightarrow \infty$ while $a_A/\Omega_A =0$, $\rho_{AB} \rightarrow
|\alpha|^2|00\rangle \langle 00|+|\beta|^2 |11\rangle\langle 11|$. In the limit of both  $a_A/\Omega_A$ and  $a_B/\Omega_B$ approach $\infty$,  $\rho_{AB} \rightarrow
|\alpha|^2|10\rangle \langle 10|+|\beta|^2 |01\rangle\langle 01|$. In both of these two limits, $S(\rho_{AB})$, $S(\rho_A)$, $S(\rho_B)$ and $I(A:B)$ all approach $-|\alpha|^2\log |\alpha|^2-|\beta|^2\log |\beta|^2$.


\section{Entanglement between qubits A and B}

Now we turn to the entanglement between qubits A and B, which are in the mixed state $\rho_{AB}$. For the two-qubit mixed state $\rho_{AB}$, a  measure  of the entanglement is the logarithmic negativity
\begin{equation}
\log ||\rho_{AB}^{T_A}||,
\end{equation}
where $||\rho_{AB}^{T_A}||$ is the trace norm of $\rho_{AB}^{T_A}$,   which is the partial transpose of  $\rho_{AB}$~\cite{vidal,plenio}. Another entanglement measure is the concurrence
\begin{equation}
C ( \rho   ) = \max  \{ {0,{\lambda _1} - {\lambda _2} - {\lambda _3} - {\lambda _4}}  \},
\end{equation}
where $\lambda_i$ ($i=1,2,3,4$) are decreasingly  ordered  eigenvalues of the matrix $\sqrt {\sqrt{\rho_{AB}}\tilde{\rho_{AB}}\sqrt{\rho_{AB}}} $, with
$
\tilde{\rho_{AB}} = (\sigma_y\otimes\sigma _y )\rho_{AB}^* (\sigma_y\otimes\sigma_y ),
$
$\sigma_y$ being the $y$-component Pauli matrix~\cite{wootters}.

\begin{figure}
\includegraphics[width=3.2in]{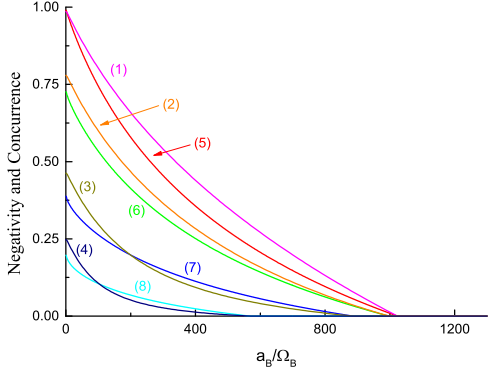}
\caption{\label{nc1}  Logarithmic negativity [plots (1) to (4)] and concurrence [plots (5) to (8)]  as functions of the acceleration-frequency ratio of qubit $B$, in the case that qubit A moves uniformly, for different initial state. (1,5): $\alpha=1/\sqrt{2}$, (2,6): $\alpha=0.4$, (3,7): $\alpha=0.2$, (4,8): $\alpha=0.1$. }
\end{figure}

\begin{figure}
\includegraphics[width=3.2in]{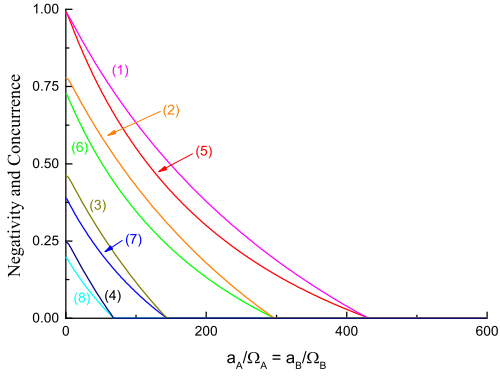}
\caption{\label{nc2}  Logarithmic negativity [plots (1) to (4)] and concurrence [plots (5) to (8)]  as functions of the acceleration-frequency ratios of $A$ and $B$, which are assumed to be the same, for different initial state. (1,5): $\alpha=1/\sqrt{2}$, (2,6): $\alpha=0.4$, (3,7): $\alpha=0.2$, (4,8): $\alpha=0.1$.}
\end{figure}

\begin{figure}
\includegraphics[width=3.4in]{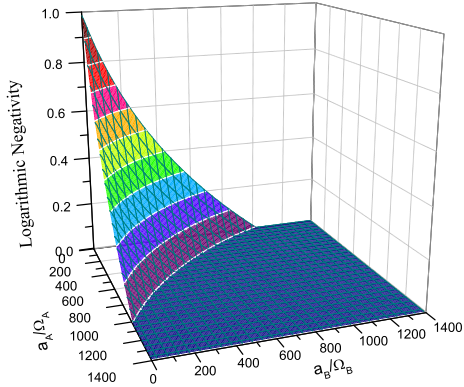}
\caption{\label{ln3d} Logarithmic negativity as a function of  $a_A/\Omega_A$ and $a_B/\Omega_B$. $\alpha=\beta=\frac{1}{\sqrt{2}}$.}
\end{figure}

\begin{figure}
\includegraphics[width=3.4in]{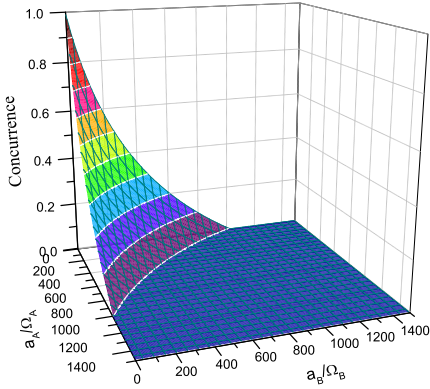}
\caption{\label{c3d} Concurrence as a function of the acceleration-frequency ratios of $A$ and $B$. $\alpha=\beta=\frac{1}{\sqrt{2}}$.
}
\end{figure}

As shown in Figs.~\ref{nc1} to \ref{c3d},  the logarithmic negativity and concurrence decrease with the increase of the acceleration-frequency ratio of each qubit and, especially, suddenly dies at a finite value of the acceleration-frequency ratio.  As can be seen in these figures, the  acceleration-frequency ratio of one qubit at which  the entanglement suddenly dies  decreases with the increase of that  of the other qubit. When one of them is zero, the other must be larger than some finite value.  In Figs.~\ref{ln3d} and (\ref{c3d}), the 3D plots of logarithmic negativity and concurrence are symmetric with respective to the plane $a_A/\Omega_A = a_B/\Omega_B$, as can be seen in the the expression of $\rho_{AB}$. These 3D plots also indicate that entanglement sudden death occurs on a curve of $a_A/\Omega_A$ and $a_B/\Omega_B$.

\section{Tripartite entanglement in three entangled accelerating qubits}

We have also extended our study to three accelerating qubits A, B and C, by using the formalism similar to the two-qubit case above. It is well known that there are two types of 3-qubit states, GHZ state
\begin{equation}
|GHZ\rangle_{ABC} = \frac{1}{\sqrt{2}} (|000\rangle + |111\rangle),
\end{equation}
and W state
\begin{equation}
|W\rangle_{ABC} = \frac{1}{  \sqrt{3}}  (|001\rangle + |010\rangle+|100\rangle),
\end{equation}
each representing a different type of tripartite entanglement~\cite{duer}.

We apply the above formalism to three entangled qubits A, B and C, as described by Eq.~(\ref{qubit}), each of which is locally coupled with a field $\Phi_q$, (q=A, B, C). For the reason given above, only the mode $\chi_{\Omega_q}$ needs to be considered.  In this way, for GHZ state, we obtain the density matrix of the qubits,
\begin{equation}
\begin{split}
{\rho _{ABC}(GHZ)} =& \frac{1}{2}C_A^2C_B^2C_C^2\sum\limits_{{n_A},{n_B},{n_C}} {\frac{{{e^{ - 2\pi \left( {{n_A}{\Omega _A}/{a_A} + {n_B}{\Omega _B}/{a_B} + {n_C}{\Omega _C}/{a_C}} \right)}}}}{{{Z_{{n_A}{n_B}{n_C}}}}} \cdot } \\
&\left[ {\left( {1 + \left( {{n_A} + 1} \right)\left( {{n_B} + 1} \right)\left( {{n_C} + 1} \right)|\mu _A|^2|\mu _B|^2|\mu _C|^2} \right)\left| {000} \right\rangle \left\langle {000} \right|} \right.\\
 +& \left( {1 + {n_A}{n_B}{n_C}|\mu _A|^2|\mu _B|^2|\mu _C|^2} \right)\left| {111} \right\rangle \left\langle {111} \right| + \left| {111} \right\rangle \left\langle {000} \right|\\
 +& \left| {000} \right\rangle \left\langle {111} \right| + \left( {{n_A}|\mu _A|^2 + \left( {{n_B} + 1} \right)\left( {{n_C} + 1} \right)|\mu _B|^2|\mu _C|^2} \right)\left| {100} \right\rangle \left\langle {100} \right|\\
 +& \left( {{n_B}|\mu _B|^2 + \left( {{n_A} + 1} \right)\left( {{n_C} + 1} \right)|\mu _A|^2|\mu _C|^2} \right)\left| {010} \right\rangle \left\langle {010} \right|\\
 +& \left( {{n_C}|\mu _C|^2 + \left( {{n_A} + 1} \right)\left( {{n_B} + 1} \right)|\mu _A|^2|\mu _B|^2} \right)\left| {001} \right\rangle \left\langle {001} \right|\\
 +& \left( {{n_A}{n_B}|\mu _A|^2|\mu _B|^2 + \left( {{n_C} + 1} \right)|\mu _C|^2} \right)\left| {110} \right\rangle \left\langle {110} \right|\\
 +& \left( {{n_A}{n_C}|\mu _A|^2|\mu _C|^2 + \left( {{n_B} + 1} \right)|\mu _B|^2} \right)\left| {101} \right\rangle \left\langle {101} \right|\\
+&\left. { \left( {{n_B}{n_C}|\mu _B|^2|\mu _C|^2 + \left( {{n_A} + 1} \right)|\mu _A|^2} \right)\left| {011} \right\rangle \left\langle {011} \right|} \right],
\end{split}
\end{equation}
where
\begin{equation}
\begin{split}
{Z_{{n_A}{n_B}{n_C}}} =& 2 +  (2{n_A}+1)|\mu _A|^2 + (2 {n_B}+1)|\mu _B|^2 + (2{n_C}+1)|\mu _C|^2 \\&+ (2 {n_A}{n_B}+n_A+n_B+1) |\mu _A|^2|\mu _B|^2 +(2 {n_A}{n_C}+n_A+n_C+1) |\mu _A|^2|\mu _C|^2 \\
 &+(2 {n_B}{n_C}+n_B+n_C+1) |\mu _B|^2|\mu _C|^2\\
 &
 + (2{n_A}{n_B}{n_C}+n_An_B+n_Bn_C+n_An_C+n_A+n_B+n_C+1) |\mu _A|^2|\mu _B|^2|\mu _C|^2.
\end{split}
\end{equation}

If the three qubits are in  W state,  their density matrix can be obtained as

\begin{equation}
\begin{split}
{\rho _{ABC}(W)} =& C_A^2C_B^2C_C^2\sum\limits_{{n_A},{n_B},{n_C}} {\frac{{{e^{ - 2\pi \left( {{n_A}{\Omega _A}/{a_A} + {n_B}{\Omega _B}/{a_B} + {n_C}{\Omega _C}/{a_C}} \right)}}}}{{{Z_{{n_A}{n_B}{n_C}}}}}} \left[ {|001\rangle \langle 010|} \right.\\
 &+ |001\rangle \langle 100| + |100\rangle \langle 001| + |100\rangle \langle 010| + |010\rangle \langle 001| + |010\rangle \langle 100|\\
 &+ {n_B}|{\mu _B}{|^2}|011\rangle \langle 110| + {n_A}|{\mu _A}{|^2}|101\rangle \langle 110| + {n_C}|{\mu _C}{|^2}|011\rangle \langle 101|\\
 &+ {n_A}|{\mu _A}{|^2}|110\rangle \langle 101| + {n_C}|{\mu _C}{|^2}|101\rangle \langle 011| + {n_B}|{\mu _B}{|^2}|110\rangle \langle 011|\\
 &+ \left( {\left( {{n_A} + 1} \right)|{\mu _A}{|^2} + \left( {{n_B} + 1} \right)|{\mu _B}{|^2} + \left( {{n_C} + 1} \right)|{\mu _C}{|^2}} \right)|000\rangle \langle 000|\\
 &+ \left( {1 + \left( {{n_A} + 1} \right){n_C}|{\mu _A}{|^2}|{\mu _C}{|^2} + \left( {{n_B} + 1} \right){n_C}|{\mu _B}{|^2}|{\mu _C}{|^2}} \right)|001\rangle \langle 001|\\
 &+ \left( {1 + \left( {{n_A} + 1} \right){n_B}|{\mu _A}{|^2}|{\mu _B}{|^2} + {n_B}\left( {{n_C} + 1} \right)|{\mu _B}{|^2}|{\mu _C}{|^2}} \right)|010\rangle \langle 010|\\
 &+ \left( {\left( {{n_A} + 1} \right){n_B}{n_C}|{\mu _A}{|^2}|{\mu _B}{|^2}|{\mu _C}{|^2} + {n_B}|{\mu _B}{|^2} + {n_C}|{\mu _C}{|^2}} \right)|011\rangle \langle 011|\\
 &+ \left( {1 + {n_A}\left( {{n_B} + 1} \right)|{\mu _A}{|^2}|{\mu _B}{|^2} + {n_A}\left( {{n_C} + 1} \right)|{\mu _A}{|^2}|{\mu _C}{|^2}} \right)|100\rangle \langle 100|\\
 &+ \left( {{n_A}\left( {{n_B} + 1} \right){n_C}|{\mu _A}{|^2}|{\mu _B}{|^2}|{\mu _C}{|^2} + {n_A}|{\mu _A}{|^2} + {n_C}|{\mu _C}{|^2}} \right)|101\rangle \langle 101|\\
 &+ \left( {{n_A}|{\mu _A}{|^2} + {n_B}|{\mu _B}{|^2} + {n_A}{n_B}\left( {{n_C} + 1} \right)|{\mu _A}{|^2}|{\mu _B}{|^2}|{\mu _C}{|^2}} \right)|110\rangle \langle 110|\\
 &+ \left.{\left( {{n_A}{n_B}|{\mu _A}{|^2}|{\mu _B}{|^2} + {n_A}{n_C}|{\mu _A}{|^2}|{\mu _C}{|^2} + {n_B}{n_C}|{\mu _B}{|^2}|{\mu _C}{|^2}} \right)|111\rangle \langle 111|} \right],
\end{split}
\end{equation}
where
\begin{equation}
\begin{split}
{Z_{{n_A}{n_B}{n_C}}} =& 3 + (3{n_A}+1) |{\mu _A}{|^2} + (3{n_B}+1) |{\mu _B}{|^2} + (3{n_C}+1) |{\mu _C}{|^2}
\\& +(3 {n_A}{n_B}+n_A+n_B) |{\mu _A}{|^2}|{\mu _B}{|^2} +(3 {n_B}{n_C}+n_B+n_C) |{\mu _B}{|^2}|{\mu _C}{|^2}\\& + (3 {n_A}{n_C}+n_A+n_C)|{\mu _A}{|^2}|{\mu _C}{|^2} \\& + (3n_An_Bn_C+n_An_B +n_Bn_C+n_An_C) |{\mu _A}{|^2}|{\mu _B}{|^2}|{\mu _C}{|^2}.
\end{split}
\end{equation}

The tripartite entanglement is the genuine three-party entanglement that cannot be reduced to any bipartite entanglement.
We use the negativity three-tangle as the measure of the tripartite entanglement~\cite{ou},  which is defined as
\begin{equation}
\pi \equiv \frac{1}{3}(\pi_A+\pi_B+\pi_C),
\end{equation}
where
\begin{equation}
\pi_A\equiv {\cal N}_{A(BC)}^2-{\cal N}_{AB}^2- {\cal N}_{AC}^2,
\end{equation}
with
\begin{eqnarray}
{\cal N}_{A(BC)} &\equiv& ||\rho_{ABC}^{T_A}||-1, \\
{\cal N}_{AB} & \equiv  &  ||\rho_{AB}^{T_A}||-1.
\end{eqnarray}
Here $\rho_{ABC}$ is the density matrix of the three qubits, $\rho_{AB}$ is the reduced density matrix of A and B, other quantities are similarly defined.

\begin{figure}
\includegraphics[width=3.2in]{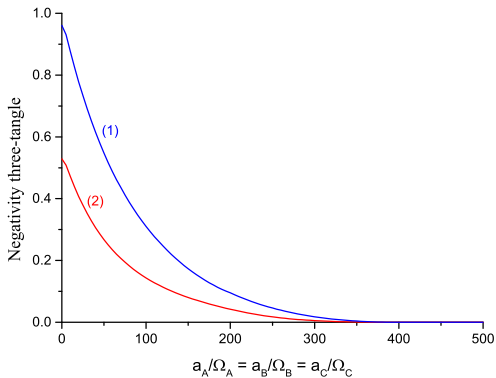}
\caption{\label{negtangle}   Negativity three-tangle  as a function of the acceleration-frequency ratio of the three qubits in  GHZ state (1) and in W state (2).}
\end{figure}

For simplicity, here we only present the result for the case that the acceleration-frequency ratios of the three qubits are the same. As shown in Fig.~\ref{negtangle}.  For either GHZ or W state,  the negativity three-tangle decreases with the increase of the acceleration-frequency ratio, and suddenly dies at a certain value. We have also found that entanglement sudden death  generally occurs when at least two qubits accelerate,  no matter whether the accelerations are equal. More details  will be discussed elsewhere.

\section{Discussion and summary}

Now we come to the black hole. The spacetime outside its horizon is described by  the Schwarzschild metric
\begin{equation}
d{s^2} =  ( {1 - \frac{{2m}}{r}}  )d{t^2} - { ( {1 - \frac{{2m}}{r}}  )^{ - 1}}d{r^2}- {r^2}d{\theta ^2} - {r^2}{\sin ^2}\theta d{\phi ^2},
\end{equation}
where the  notations are standard. The proper acceleration of a static observer at $r$ is
\begin{equation}
a = \frac{m}{{{r^2}}}{ ( {1 - \frac{{2m}}{r}}  )^{ - 1/2}}.
\end{equation}
It is well known that near the horizon $r=2m$,   the Schwarzschild metric can be approximated as the Rindler metric~\cite{rindler,eduardo}.  The closer $r$ is to the horizon, the larger the acceleration.  The uniform movement   corresponds to free falling into the black hole.

Consider  the entangled states studied above.  Suppose at most  one of them freely fall into the black hole, while the other is  near the horizon $r=2m$. According to the calculation above,  we know that the entanglement between the qubits suddenly dies when one or more accelerating qubits are close enough to the horizon. For each qubit near the horizon,  we have
\begin{equation}
r\approx 2m[1-\frac{1}{(4m a)^2}]^{-1},
\end{equation}
from which the location of the entanglement sudden death can be determined.

Finally, one may wonder the reason of the entanglement sudden death as  studied here. We think the Bosonic fields act as a drain of the entanglement originally exists between the qubits, because there are infinite number of Fock states $|n\rangle$  for each Bosonic field mode. In contrast, we conjecture that there is no entanglement sudden death if the fields are Fermionic as  there are only two Fock states $|0\rangle$ and $|1\rangle$  for each Fermionic field mode. The absence of entanglement sudden death was recently noted in entanglement between Unruh modes~\cite{richter}. Our work implicates that entanglement sudden death of the qubits can act as a probe of the nature of ambient fields coupled with the qubits and the existence of acceleration or gravity.

To summarize, we have studied the entanglement of two accelerating qubits coupled with scalar   fields, and demonstrate the occurrence of  its sudden death.  We also found the entanglement sudden death of tripartite entanglement of three accelerating qubits in GHZ and W states. These results imply the entanglement sudden death of field-coupled qubits near the horizon of a black hole.  This work might be useful to  the issue of black hole firewall~\cite{almheiri} or energetic  curtain~\cite{braunstein}.


\begin{thebibliography}{99}
\bibitem{schroedinger} E. Schr\"{o}dinger, Proc. Camb. Phi. Soc. {\bf 31}, 555 (1935).
\bibitem{fulling} S. A.  Fulling, Phvs. Rev. D {\bf 7}, 2850 (1973).
\bibitem{davies} P. C. W. Davies, J. Phys. A {\bf 8}, 609 (1975).
\bibitem{dewitt} B. S. DeWitt, Phys. Rep. {\bf 19}, 295 (1975).
\bibitem{unruh} W. G. the Unruh, Phys. Rev. D {\bf 14}, 870 (1976).
\bibitem{unruhwald} W. G. the Unruh and R. M. Wald, Phys. Rev. D {\bf 29}, 1047 (1984).
\bibitem{matsas} L. C. Crispino, A. Higuchi and G. E. Matsas, Rev.  Mod. Phys. {\bf 80}, 787 (2008).
\bibitem{mukhanov} V. F. Mukhanov and S. Winitzki, {\em Introduction to Quantum Effects in Gravity}, Cambridge University Press (2007).
\bibitem{fuentes} I. Fuentes-Schuller and R. B. Mann,  Phys. Rev. Lett. {\bf 95}, 120404 (2005).
\bibitem{eduardo} E. Marti\'{i}n-Mart\'{i}nez, L. J. Garay and J. Le\'{o}n, Phys. Rev. D {\bf 83}, 012111 (2011).
\bibitem{hwang} M.-R. Hwang, D. Park and E. Jung, Phys. Rev. A {\bf 83}, 012111 (2011).
\bibitem{shamirzai} M. Shamirzai, B. Nasr Esfahani and M. Soltani,  Int. J. Theor. Phys. {\bf 51}, 787 (2012).
\bibitem{alsing} P. M. Alsing, I. Fuentes-Schuller, R. B. Mann and T. E. Tessier, Phys. Rev. {\bf A 74}, 032326 (2006).
\bibitem{jing1} Q. Pan and J. Jing, Phys. Rev. A {\bf 77}, 024302 (2008).
\bibitem{jing2} J. Wang and J. Jing, Phys. Rev. A {\bf 83}, 022314 (2011).
\bibitem{richter} B. Richter and Y. Omar, Phys. Rev. A {\bf 92}, 022334 (2015).
\bibitem{shi} Y. Shi, Phys. Rev. D {\bf 70}, 105001 (2004).
\bibitem{yu} T. Yu and J. H. Eberly, Science 323, 598 (2009).
\bibitem{hu0} S.-Y. Lin, C.-H. Chou and B. L. Hu, Phys. Rev. D. {\bf 78}, 125025 (2008).
\bibitem{hu1} B. L. Hu, S.-Y. Lin and J. Louko,  Class. Quantum Grav. {\bf 29}, 224005 (2012).
\bibitem{hu2} R. Zhou, R. O. Behunin, S.-Y. Lin and B. L. Hu, JHEP {\bf 07}, 072 (2013).
\bibitem{matsas1} A. G. S. Landulfo and G. E. A. Matsas, Phys. Rev. A {\bf 80}, 032315 (2009).
\bibitem{doukas1} J. Doukas and L. C. L. Hollenberg, Phys. Rev. A {\bf 79}, 052109 (2009).
\bibitem{doukas2} J. Doukas and B. Carson, Phys. Rev. A 81, 062320 (2010).
\bibitem{huyu} J. Hu and H. Yu, Phys. Rev. A {\bf 91}, 012327 (2015).
\bibitem{datta} A. Datta, Phys. Rev. A {\bf 80}, 052304 (2009).
\bibitem{celeri} L. C. C\'{e}leri, A. G. S. Landulfo, R. M. Serra and G. E. A. Matsas, Phys. Rev. A {\bf 81}, 062130 (2010).
\bibitem{yao} Y. Yao, X. Xiao, L. Ge, X.-G. Wang and C.-P. Sun,
Phys. Rev. A {\bf 89}, 042336 (2014).
\bibitem{wang} J. Wang, Z. Tian, J. Jing and H. Fan, Sci. Rept. {\bf 4}, 7195 (2014).
\bibitem{kok} P. Kok and U. Yurtsever, Phys. Rev. D {\bf 68}, 085006 (2003).
\bibitem{vidal} G. Vedal and R. F. Werner, Phys. Rev. A {\bf 65}, 032314 (2002).
\bibitem{plenio} M. B. Plenio, Phys. Rev. Lett. {\bf 95}, 090503 (2005).
\bibitem{wootters} W. Wootters, Phys. Rev. Lett. {\bf 80}, 2245 (1998).
\bibitem{duer} W. D\"{u}r, G. Vidal and J. I. Cirac, Phys. Rev. A {\bf 62}, 062314 (2000).
\bibitem{ou} Y. C. Ou and H. Fan, Phys. Rev. A {\bf 75}, 062308 (2007).
\bibitem{rindler} W. Rindler, Am. J. Phys. {\bf 34}, 1174 (1966).
\bibitem{almheiri} A. Almheiri, D. Marolf, J. Polchinski and J. Sully, JHEP {\bf 02}, 062 (2013).
\bibitem{braunstein} S. L. Braunstein, S.
Pirandola and K. Zyczkowski,  Phys. Rev. Lett. {\bf 110}, 101301
(2013).
\end{thebibliography}
\end{document}